\documentclass[prl,aps]{revtex4}
\usepackage{bm}
\usepackage{setspace}
\usepackage{graphicx}
\usepackage{amsfonts}
\usepackage{amsmath}
\usepackage{amssymb}
\newcommand {\be}{\begin{equation}}
\newcommand {\ee}{\end{equation}}
\newcommand {\bea}{\begin{eqnarray}}
\newcommand {\eea}{\end{eqnarray}}

\begin{document}

\title{ Analytical solution for optimal squeezing of wave packet of a trapped quantum particle}
\author{Ilya Grigorenko}
\affiliation{Theoretical Division T-11, Center for Nonlinear
Studies, Center for Integrated Nanotechnologies, Los Alamos
National Laboratory, Los Alamos, New Mexico 87545, USA}

\date{\today}

\begin{abstract}
Optimal control problem with a goal to squeeze  wave packet of a
trapped quantum particle is considered and solved analytically
using adiabatic approximation. The analytical solution that drives
the particle into a highly localized final state is presented for
a case of an infinite well trapping potential. The presented
solution may be applied to increase the resolution of atom
lithography.
\end{abstract}

\maketitle 

 {
 The recent interest in squeezed quantum states
has stimulated  research of optimal squeezing of quantum wave
packets. Several schemes, including numerical solution of optimal
control problem \cite{optimal_squeezing,optimal_control}, or
iterative pump-dump technique by switching the wavefunction
between the ground and excited states with $\pi$ pulses
\cite{pump_dump_squeezing}, were introduced. Another method, based
on classical parametric squeezing by a sudden increase of the
potential depth was considered  for
the case of 3D optical lattice gases \cite{parametric_driving_squeezing}. 
Other, more efficient methods, need relatively short $\delta$-like
control pulses \cite{leibscher_averbukh}. An experimental
realization of this squeezing algorithm using impulse periodic
potential  \cite{raizen} demonstrated the fundamental limitation for
squeezing due to the nonzero duration of the kicking pulses. An
interesting analytical solution was proposed for controls on time
scales shorter than classical Kepler orbit period \cite{analyt}.
However, most of the above mentioned methods were based on numerical
solutions and have demonstrated limited success. }

{ One of the significant applications of squeezed atomic wave
packets is atom lithography \cite{555}. Atom lithography is
relatively cheap and powerful method that does not require any
material mask to fabricate semiconductor, metallic or magnetic
nanostructures. The method is based on manipulation of atoms by
laser fields on a scale under $100$ nm, and  it is considered as a
promising way to overcome the resolution limits of usual optical
lithography \cite{leibscher_averbukh}. If we know that at a moment
$T$ a cold atom will arrive at a surface, then the uncertainty of
its position in the surface's plane may be roughly described in
terms of the corresponding width of the atomic wave packet (we
neglect diffusion on the surface and other effects that also have
contribution to the uncertainty of the final atomic position).

 Another possible application of squeezed molecular wave packets
is mapping of the potential surfaces \cite{map}. Using
"pump-probe'' measurement technique the excited state dynamics can
be followed and the potential surface can be mapped.   The ability
to map potential surfaces using such techniques requires
generation of very localized wave packets
\cite{optimal_squeezing}.

 In this paper we are going to consider the problem of squeezing
of a  wave packet of a  trapped quantum particle. Using our
strategy one can obtain, in principle, an arbitrary narrow wave
packet, that, in return, results, for example, in ultrahigh
resolution of atomic lithography. However, it is clear that the
squeezing of wave packets is limited by the Heisenberg principle.
For example, a Gaussian-like wave packet with the width $\sigma$
has the kinetic energy $E_{kin}\propto\sigma^{-2}$. Thus, one
needs to provide the controlled system with enough amount of
energy. The amplitude of the control field should be relatively
big in order to achieve wave packet localization. As a result,
perturbation theory may not be applicable, and numerical solution
of the time-dependent problem may be necessary. However, we are
going to show that under curtain approximations the optimal
control field and the corresponding wave packet evolution can be
described analytically.

 Let us consider a simplified $1$D picture, where  a quantum
particle of mass $m$ is trapped in the  potential $U_{trap}(x)$
and interacts with a time-dependent control potential
$U_{c}(x,t)$. Since in this work optimal control is achieved using
linear polarized field and therefore can be performed in $x$ and
$y$ directions independently, 2D generalization of the presented
approach is straightforward.
 In the absence of $U_{c}(x,t)$, the eigenstates
 and the eigenenergies of the trapped particle  are $\psi_j(x)$
and $E_j$ ($j=1,2,...$) correspondingly.

 The dynamics of  the particle's wave packet is described by the
time-dependent Schr\"odinger equation
\begin{equation}
\label{eq1}
 i\hbar\frac{\partial \Psi}{\partial t}=(H_0+U_c(x,t))\Psi,
\end{equation}
where $H_0=\frac{\hbar^2}{2m}\frac{d^2}{dx^2}+U_{trap}(x)$ is the
unperturbed Hamiltonian. We assume that initially  system is in
the ground state $\Psi(x,0)\equiv\psi_1(x)$. Note,  we neglect in
this work any kind of decoherence effects (including spontaneous
emission). If $T$ is the duration of the control time interval and
$\gamma$ is the characteristic decoherence rate in the system, we
assume that $\gamma T \ll 1$. 
We consider the  control potential in the form:
\begin{equation}
\label{field_sum} U_c(x,t)= U_{in}(x)\sum_{j=2}^{+\infty}
V_j(t)\cos(\omega_{1j}t ),
\end{equation}
where   $\omega_{1j}=(E_j-E_1)/\hbar$ is the transition frequency
between levels $1$ and $j$, and $V_j(t)$ is the envelop of the
component of the control field with the carrier frequency
$\omega_{1j}$. Our guess of the time dependent structure of the
optimal control potential $U_c(x,t)$ is based on a natural
assumption that the most efficient control is achieved within the
resonance coupling \cite{akulin}. The function $U_{in}(x)$
characterizes the inhomogeneity of the control potential. As an
example, one can consider form $U_{in}(x)\propto \sin(\beta
x)/\beta$ one and $U_{in}(x)\propto x$ represents the long
wavelength limit $\beta\to0$ (linear potential, constant force).
This potential can be realized in different ways. For example, one
can use locally inhomogeneous electric field, so the force acting
on a neutral atom is proportional to the gradient of the field
intensity \cite{cohen_tannuji}. Another approach is to use ions
instead of neutral atoms to control. In this case the force acting
on an ion is simply proportional to its charge and the local
amplitude of the applied field.

The optimal control problem is usually formulated as the
following: starting from the initial state $\psi_1(x)$, one would
like to steer the system optimally to a target state
$\Psi_T(x)\equiv\Psi(x,T)$ at time $t=T$, such that the spatial
dispersion $\sigma$
of the wave packet is minimal: 
$\sigma^2=<\Psi_T|x^2|\Psi_T>-<\Psi_T|x|\Psi_T>^2\to min.$
Obviously, this condition on $\Psi_T$ is vague, because there is an
infinite amount of wave packets with the same $\sigma$, and as a
result, the numerical algorithm can be easily trapped in local
minima. In other words, the condition $\sigma\to min$ allows very
large search space, that may complicate iterative numerical solution
of the optimal control problem (see for example,
\cite{optimal_squeezing}).

 A simple way to handle this uncertainty is to determine a target
wave packet with a {\it specified shape}, that can be parameterized
by $\sigma$. Thus, one can formulate optimal control problem that
has a unique solution for arbitrary small $\sigma$. The problem is
then split into two parts, and the first part is to find an
appropriate shape of the target wave packet. A reasonable choice can
be made if we look for a strongly squeezed wave packet with a
minimum kinetic energy for a given width. In this case we minimize
the necessary work to be done to build up such a wave packet. It is
easy to show, that the ground state of a harmonic potential
satisfies this criterium. However, as we show below, in the limit of
long wavelength control field the initial and the target wave
functions should have different symmetry, i.e. if we start from the
ground state, which is obviously a symmetric (even) function, one
needs an antisymmetric (odd) target wave function. This is the
reason to choose the target wave function in the form of the first
excited state of a harmonic potential.


 We search for a control field that is optimal in the sense of
spending minimum amount of energy that is still enough to reach the
target state. Thus, we introduce a constraint that effectively
limits the total field energy
$\sum_{j=2}^{+\infty}\int_0^T{|V_j(t)|^2 dt}=E_{tot}$. Note, that
without this constraint optimal control problem has no finite
solution, because  as we mentioned above, one needs an infinite
amount of energy $E_{tot}\to +\infty$ to squeeze a broad wave packet
in order to achieve $\sigma\to 0$.

 Using expansion in a real eigenbasis of the trapping potential
$\Psi(x,t)=\sum_{j=1}^{+\infty} a_j(t) \exp(-i E_j t
/\hbar)\psi_j(x)$, Eq.(\ref{eq1}) can be written as

\begin{eqnarray}
\label{basis} i\hbar\dot{a}_j=\sum_{p=2}^{+\infty}
V_p(t)\cos(\omega_{1p} t)\sum_{k=1}^{+\infty}a_k(t)e^{-i\frac{ E_k
t}{\hbar}} d_{kj},
\end{eqnarray}
where  $d_{kj}$ is the coupling matrix element between states $k$
and $j$: $d_{kj}=\int_{-\infty}^{+\infty}\psi_k(x) U_{in}(x)
\psi_j(x)dx$.

The  optimal control problem is to find the time dependent field
envelopes $V_j(t)$ that drive the system to the final state
 characterized by the set of amplitudes $\{a_j^T\}$ at time $T$,
$a^T_j=\int_{-\infty}^{+\infty}\psi_j(x) \Psi_T(x)dx$. The
 Lagrangian of this optimal control is \cite{optimalcontrol}:
\begin{equation}
\label{lagrange}
 L=\int_0^T \sum_{j=1}^{+\infty}(
|a_j(t)-a_j^T|^2\delta(t-T)+\lambda  V_j^2(t))dt,
\end{equation}
where $\lambda$ is a Lagrange multiplier and $\delta(t)$ is the
Dirac delta function. The problem of optimization of the
Lagrangian Eq.(\ref{lagrange}) subject to non-holonomic
constraints given by Eq.(\ref{basis}) is very complicated, and
needs some simplifications.

 Let us assume the adiabatically slow changes of the field
 envelopes $V_j(t)$ on the characteristic time scale
 $\max\{\omega_{1j}^{-1}\}$. Applying the Rotating Wave
 Approximation (RWA)\cite{RWA} to Eq.(\ref{basis}) one gets:
\begin{eqnarray}
\label{res}
i\hbar\dot{a}_j=\frac{1}{2}\sum_{k,p}V_p(t)(\Delta(\hbar\omega_{1p}-E_k+E_j)\nonumber\\+\Delta(-\hbar\omega_{1p}-E_k+E_j))
a_k(t)d_{kj},
\end{eqnarray}
where the  function $\Delta()$ returns unity if its argument
equals to zero, and zero in all other cases. Thus, all the
off-resonance quickly oscillating terms are set to zero. In
Eq.(\ref{res}) we account all resonance transitions, however, as
we see later, we can neglect transitions between initially
unoccupied states.

Now we make a crucial approximation that significantly reduces the
complexity of the problem, and it is based on the assumption that
the target wave packet is narrow.
  If $\sigma$ is small, $\sigma/L \ll 1$ ($L$ is the
characteristic length scale of the trapping potential
$U_{trap}(x)$), it is easy to show that the target amplitudes
$a^T_j$  are also small,  $|a^T_j|\propto\sqrt{\sigma/L}\ll1$. It
is natural that the overlap between an eigenfunction $\psi_j$ and
a narrow target wave packet is small.

Since under the optimal control the amplitude of an excited state
$a_j(t), j>1$ should grow {\it monotonously} from its zero initial
value to its final value $|a^T_j|\ll1$, one can simply neglect
$a_j(t), j>1$ compare with $a_1(t)$. Thus, Eq.(\ref{res}) can be
simplified, since the $|a_1(t)| \gg  |a_j(t)|$ most of the time,
except may be for a short time interval close to $T$. After these
simplifications Eq.(\ref{res}) becomes:
\begin{eqnarray}
\label{simplified} i\hbar\dot{a}_1=\frac{1}{2}\sum_{k=2}^N V_k(t)a_k(t)d_{k1},\\
i\hbar\dot{a}_k= \frac{1}{2}V_k(t)a_1(t)d_{k1}.\nonumber
\end{eqnarray}
We have performed comparison between  numerical solutions of
Eq.(\ref{res}) and Eq.(\ref{simplified}), and for small parameter
values $\sigma\le 10^{-2} L$ found a good agreement ($<1\%$).

 Using vector representation of Eq.(\ref{simplified}):
$\dot{\vec{a}}=\hat{Z}(t)\vec{a}$ and the Magnus series expansion
\cite{magnus}, we can write the exact solution as
\begin{equation}
\label{magn} \vec{a}(t)=e^{\int_0^t{\hat
Z}(s_1)ds_1+\frac{1}{2}\int_0^t\Big[{\hat
Z}(s_1),\int_0^{s_1}{\hat Z}(s_2)ds_2 \Big]ds_1+...}\vec{a}(0),
\end{equation}
where vector $\vec{a}(0)$ determines the initial wave packet
$\Psi(x,0)$, and $[\:,\:]$ denotes the commutator.

We neglect all the terms in the series Eq.(\ref{magn}) except the
first one. It is
 a good approximation, if the first term in the series is much
larger than the second one, and all further terms. Using the
explicit form of the operator ${\hat Z}(t)$ and the Mean Value
Theorem, we obtain $\big|V(t)\big|\gg T^2 \big|\frac{\partial
V(t_1)}{\partial t}\big|^2$, where $t,t_1\in(0,T)$. This is the
adiabatical condition, under which it is possible to integrate
Eq.(\ref{simplified}) analytically. Introducing new variables
$\theta_k(t)=1/2\int_0^t V_k(t^\prime)dt^\prime$, and
$R(t)=\Big(\sum_{k=2}^{+\infty} \theta_k^2(t) d_{k1}^2\Big)^{1/2}
$, one can write an approximate solution given by Eq.(\ref{magn})
as:
\begin{eqnarray}
\label{occupations} a_1(t)=\cos\Big(R(t)/\hbar\Big),\nonumber\\
a_j(t)= d_{j1} \theta_j(t) \sin\Big (R(t) /\hbar\Big)/R(t),
j=2,3,...
\end{eqnarray}
We can use this solution to substitute into the Lagrangian
Eq.(\ref{lagrange}), that  allows us to derive explicit
Euler-Lagrange equations with respect to unknown $\theta_k(t)$.
Note, that the approximations made in Eqs.(\ref{res},\ref{magn})
do not affect the normalization condition: $\sum_{k=1}^{+\infty}
|a_k(t)|^2=1$, unlike if one uses the simple first order
perturbation theory.

The Euler-Lagrange equations which determine optimal
$\theta_k(t)$, are:
\begin{eqnarray} \label{2f}
 \lambda \frac{d^2}{d t^2}\theta_k- \frac{\partial }{\partial
 \theta_k}\sum_j |a_j(\{\theta_k\})-a_j^T|^2\delta(t-T)=0,
\end{eqnarray}
where we use the explicit notation $a_j(\{\theta_k\})$ to stress
the dependence of amplitudes on the set of unknown functions
$\{\theta_k\}$.

Since in practice the target state has a finite (although may be
very small) width $\sigma$, it is reasonable to keep a limited
number of energy levels $N$ of the controlled system in the
consideration. For a smaller $\sigma$ one needs to increase $N$.
 The appropriate number $N$ can be determined from the condition
$\sum_{i=N+1}^{+\infty}  |a_i^T|^2<C_{const}\ll1$.

Note, that under the discussed approximations optimal control
problem Eq.(\ref{2f}) in new variables becomes particular simple and
formally splits into $N-1$ {\it independent} control problems for
the whole control interval (since the delta function takes zero
value), except at the time $t=T$. Each of the equations is
equivalent to find an optimal envelope $V_k(t)$ for one resonant
transition.

Integration of Eq.(\ref{2f}) gives $\theta_k =A_k+B_kt$. This result
is obtained under the condition that $\lambda\ne0$. Thus, linear
dependence of $\theta_k$ on time is a consequence of the constraint
on the energy, introduced in Eq.(\ref{lagrange}). $A_k$ can be
readily determined from the initial condition $\theta_k(0)=0$, that
gives  $A_k=0$. The coefficients $B_k$ are determined from the
condition that Eq.~(\ref{2f}) should also be satisfied at $t=T$.
This is equivalent that the second term in Eq.~(\ref{2f}) should
turn exactly to zero at $t=T$. This gives us a system of algebraic
equations for $B_k$:
\begin{eqnarray}
\label{bound} a_j(\{\theta_k\})|_{t=T}=d_{j1} B_j  T  \sin\Big (R(T)
/\hbar\Big)/R(T)=a_j^T,j=2,3,...
\end{eqnarray}
with $R(T)=T \Big(\sum_{k=2}^{+\infty} B_k^2 d_{k1}^2\Big)^{1/2}$.
Eq.~(\ref{bound}) actually plays a role of the second boundary
condition for each $\theta_k(t)$ at $t=T$. It is easy to verify that
the solution (see solution of a similar problem in
\cite{optimalcontrol})
\begin{eqnarray}
\label{solution} \dot{\theta}_k(t)=B_k = \frac{1}{2}
V_k(t)=\frac{\hbar\pi a^T_k}{2 d_{k1} T}, k=2,3,...
\end{eqnarray}
satisfies Eq.~(\ref{bound}). Here we have assumed that the symmetry
of the target wave packet is chosen such, that if $d_{k1}=0$ then
$a^T_k=0$, and the correspondent $V_k(t)$ is also zero.

The result Eq.(\ref{solution}) self-consistently  justifies the
adiabatical approximation made for the Magnus expansion
Eq.(\ref{magn}), since for the constant field envelopes all terms
except the first one in Eq.(\ref{magn}) are equal to zero.
Substituting Eq.(\ref {solution}) into Eq.(\ref{occupations}) one
easily gets the corresponding dynamics for the occupation numbers
$a_k(t)$. Using the obtained solution one also can estimate that the
approximation made for Eq.(\ref{res}) is a good one for times
$t\in[0,T-2T\sqrt{\sigma/L}/\pi)$.

 The cost of all the simplifying assumptions (including RWA), which lead to  Eq. (\ref{2f}), is
 that Eq.~(\ref{2f}) is a system of the second order differential
equations for unknown $\theta_k$, and one can not impose more than
two boundary conditions for each equation. The constraint on the
energy can be seen as a third condition, but for the whole system.
By substituting into the energy constraint $\sum_{k=2}^\infty
\int_0^T |V_k|^2dt=E_{tot}$, one obtains $\int_0^T
\sum_{k=2}^\infty|{\dot{\theta}_k}|^2dt=\sum_{k=2}^\infty|B_k|^2
T=E_{tot}$. Since $B_k$ are already determined from the condition at
$t=T$, the energy constraint $E_{tot}$ cannot be arbitrary. However,
one can assume that duration of the control $T$ is unknown at the
beginning, so one can determine the control time $T$ through a given
total energy of the control field $E_{tot}$.

Now let us consider an example of a trapped quantum particle in an
infinite potential well of width $L$, interacting with a control
potential $U_c(x,t)$ in the long wavelength limit $\beta\to 0$.
\begin{figure}
\vspace{0.cm}
\includegraphics[width=7.cm,angle=0]{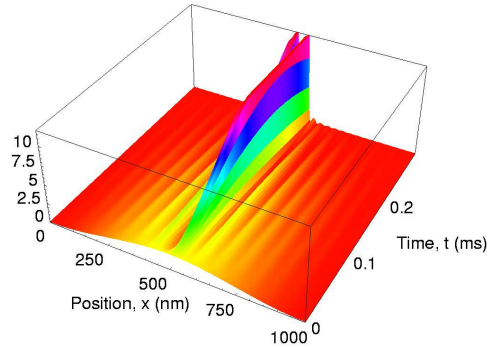}
\caption{\label{fig1} \footnotesize{ (Color online). Time evolution
of the controlled wave packet density $|\Psi(x,t)|^2$ in an infinite
rectangular well. Note monotonic increase of the wave packet height
and decrease of its width. }}
\end{figure}
As we mentioned above, it is easy to show that the matrix element
$d_{j1}=<\psi_k|U_{in}|\psi_1>=-4 L(\cos(\pi j)j+j)/(\pi^2(1-2 j^2
+j^4))$, so $d_{j1}=0$, if $j$ is odd, and $d_{j1}\ne 0$ if $j$ is
even. We assume the system initially is in the ground state, thus
it is impossible to drive the system into a squeezed target state
with the same symmetry in the long wavelength limit.
 In order to utilize the only allowed ground state-odd states
transitions, the target state $\Psi_T(x)$ is chosen to be an
antisymmetric  wave packet
  parameterized by $\sigma$ and located at  position $x_0$:
\begin{eqnarray}
\label{target} \Psi_T(x)=B(x-x_0)\exp(-(x-x_0)^2/\sigma^2),
\end{eqnarray}
where the constant $B$ is determined from the normalization
condition over the interval $[0,L]$. In the limit $\sigma\to 0$,
$\Psi_T(x)$ becomes equivalent to the first derivative of the
Dirac delta function $\delta^\prime(x-x_0)$, and it has the
minimum kinetic energy among all other wave packets with the same
symmetry and given width $\sigma$.
 The amplitudes of the final state in the basis of the infinite
well potential $a^T_j$ can be calculated analytically in terms of
the Erf(x) function.

The ultimate goal of a useful squeezing algorithm is to create a
wave packet with a characteristic size much less than the
characteristic length scale of the inhomogeneous control potential.
A possible realization of the proposed squeezing method in the long
wave length limit is to control ions of charge $+q$ in the
oscillating linear potentials. Let us assume for simplicity $q=|e|$.
In a typical experiment let us assume the intensity of the highest
mode of the controlling electromagnetic field is of the order of
$100$ mW/cm$^2$, that corresponds to the electric field of
$E=8.0\times10^{-2}$V/m. We assume the initial width of the wave
packet to be of the order of the infinite well trapping potential
$L=1\times10^{-6}$m, and the final width of the wave packet
$\sigma\approx L/50=20$ nm. Using Eq.(\ref{solution}), we estimate
duration of the squeezing procedure $T=\pi\hbar a_{30}^T/(E
d_{1\;30})\approx 3\times10^{-4}$ s.

 In Fig. 1 we show time evolution of  the wave packet density
 $|\Psi(x,t)|^2$, trapped in the infinite well potential
 $x\in(0,L)$ under the control of the optimal field determined by
 Eq.(\ref{solution}). We choose $N=30$ of controlled levels, $x_0=L/2$, and
 final width $\sigma=L/50$.
Note, that the height of the controlled wave packet  is increasing
monotonically, while the width is monotonically decreasing to its
theoretical target limit. This is a direct consequence that the
control is optimal, opposite to the previous studies (see, for
example, Fig. 3 in \cite{optimal_squeezing}, the non-monotonous
decrease of the width of the wave packet is an indication that the
numerical solution is a local extremum).

If it is more preferable to use neutral atoms instead of ions, then
the necessary control potential can be created using dipole forces
which are proportional to the amplitude gradient of the laser field
\cite{cohen_tannuji}. Note, that unlike in the case of ions, the
dipole force arises for sufficiently inhomogeneous laser fields.

Another possible way to realize the proposed squeezing method is to
consider control of molecular wave packets, or electron wave packets
in artificial molecules: quantum dots. In this case one starts from
the ground state and finishes at the linear combination of the
vibrational (or eigenstates of the confining potential in the case
of quantum dots) states, that builds up a narrow wave packet at a
given time $T$. The latter can be probed by means of the ultrafast
spectroscopy.

In this work we have derived {\it analytically} the control
potential to localize an atomic wave packet in real space. This
method is based on our knowledge of the eigenenergies and
eigenfunctions of the controlled system in the trapping potential,
and it is rigorous under the approximations made. Note, we can
achieve the wave packet squeezing using control potentials with
much larger wavelength ($\beta \to 0$) compare to the
characteristic scale of the squeezed wave packet $\sigma$, that
may be useful in atom lithography.

One essential assumption we made is a  strong unharmonicity of the
trapping potential. This assumption assures us that resonance
frequencies between the initial (ground) state and multiple final
states are all well separated. The trapping potential can be
different from the infinite well potential considered in this work,
but it will be more difficult to obtain solution  in a simple
analytical form. The presented  method breaks down for relatively
short control intervals $T$ or relatively strong control fields,
when the RWA is not applicable. Another limitation is that the
control interval cannot be very large, since the approximation that
decoherence effects are negligible $T\gamma\ll1$ should hold.

The resulting optimal control field has a complicated spectrum and
shape, and it may be hard to obtain it using the standard
numerical solution techniques for optimal control problems
\cite{optimal_squeezing}. That makes analytical form of the
presented solutions even more attractive.





This work was carried out under the auspices of the National Nuclear
Security Administration of the U.S. Department of Energy at Los
Alamos National Laboratory under Contract No. DE-AC52-06NA25396. }

\end{document}